\documentclass[aps,floatfix,twocolumn,superscriptaddress,amsmath,amssymb,showpacs,10pt]{revtex4-1}
\usepackage{graphicx}
\usepackage{epstopdf}
\usepackage{epsfig}
\usepackage{epsf}
\usepackage{url}
\usepackage[USenglish]{babel}
\usepackage{hyperref}
\edef\SupplementaryMultimediaURL{https://doi.org/10.6084/m9.figshare.32812943}
\newcommand{\SupplementaryMultimediaBibLink}{\href{\SupplementaryMultimediaURL}{\url{\SupplementaryMultimediaURL}}}
\newcommand{\im}{\mathrm{i}}
\def\bcen{\begin{center}}
\def\ecen{\end{center}}
\allowdisplaybreaks
\usepackage{verbatim}
\usepackage{natbib}
\usepackage{amsmath, nccmath}
\usepackage[export]{adjustbox}
\usepackage{dcolumn}
\usepackage{bm}
\usepackage{bbm}
\usepackage{lipsum}
\usepackage{color}
\usepackage{newtxtext}
\usepackage{newtxmath}
\usepackage{sidecap}
\usepackage{tikzit}

\tikzstyle{qtt}=[fill=white, draw=black, shape=rectangle, line width=0.75pt]
\tikzstyle{leftorth}=[fill=white, draw=black, shape=regular polygon, regular polygon sides=3, shape border rotate=270, line width=0.75pt, scale=0.5]
\tikzstyle{rightorth}=[fill=white, draw=black, shape=regular polygon, regular polygon sides=3, shape border rotate=90, line width=0.75pt, scale=0.5]
\tikzstyle{diamond}=[fill=white, draw=black, shape=diamond, line width=0.75pt, scale=0.6]
\tikzstyle{diamondinv}=[fill={black!30}, draw=black, shape=diamond, line width=0.75pt, scale=0.6]
\tikzstyle{blackqtt}=[fill=black, draw=black, shape=rectangle, line width=0.5pt]
\tikzstyle{2-leg tensor}=[fill=white, draw=black, shape=rectangle, minimum width=0.5cm, line width=0.75pt]
\tikzstyle{2-leg tensor round}=[fill=white, draw=black, shape=rectangle, minimum width=0.5cm, line width=0.75pt, rounded corners=3pt, tikzit shape=circle]
\tikzstyle{circle}=[fill=white, draw=black, shape=circle, line width=0.75pt, scale=0.75]
\tikzstyle{blackcircle}=[fill=black, draw=black, shape=circle, line width=0.75pt, scale=0.75]
\tikzstyle{circleinv}=[fill={black!30}, draw=black, shape=circle, line width=0.75pt, scale=0.75]
\tikzstyle{smallcircle}=[fill=white, draw=black, shape=circle, line width=0.75pt, scale=0.5]
\tikzstyle{onehot}=[fill=black, draw=black, shape=circle, line width=0.75pt, scale=0.3]
\tikzstyle{hexagon}=[draw, fill=white, regular polygon, regular polygon sides=6, minimum size=1cm, line width=0.75pt, scale=0.28]
\tikzstyle{pentagon}=[draw, fill=white, regular polygon, regular polygon sides=5, minimum size=1cm, line width=0.75pt, scale=0.28]

\tikzstyle{normal}=[-, line width=0.75pt]
\tikzstyle{dashed}=[-, dashed]

\input{qtt.tikzdefs}
\usepackage[section]{placeins}
\usepackage{tikz}
\usepackage{tikz-feynman}
\tikzfeynmanset{warn luatex=false, compat=1.1.0}
\tikzfeynmanset{
  every boson/.style={
    draw=black,
    decorate,
    decoration={coil, aspect=0.6, segment length=4mm, amplitude=2mm},
    line width=1.2pt
  },
  every fermion/.style={
    line width=1.2pt
  }
}

\begin{document}
	
	\title{Nonequilibrium electron-phonon dynamics with high momentum resolution:\\ Thermalization bottlenecks and 
	the effects of phonon dispersion}
	
	\author{Maksymilian \'Sroda}
	\affiliation{Department of Physics, University of Fribourg, 1700 Fribourg, Switzerland}
	\author{Philipp Werner}
	\affiliation{Department of Physics, University of Fribourg, 1700 Fribourg, Switzerland}

\begin{abstract}
The nonequilibrium interplay of electrons and phonons plays an important role in the thermalization of solids, yet the microscopic picture of transient states and relaxation pathways remains incomplete. 
Previous nonequilibrium Green's function (NEGF) studies with full two-time dependence were restricted to local phonons and local self-energy approximations, leaving momentum-dependent phonon dynamics largely unexplored within such an accurate approach.
In this work, we demonstrate the power of the recently developed quantics-tensor-train (QTT) NEGF framework with large-scale lattice simulations of models with arbitrary phonon dispersions.
QTTs provide a memory-efficient representation of the full two-time Green's functions, enabling momentum-resolved simulations with full electron-phonon feedback on lattices up to $256\times 256$ sites.
By systematically comparing optical and acoustic phonon models, we reveal a hierarchy of relaxation bottlenecks that substantially extends the well-known phonon-window bottleneck effect.
For optical phonons, we confirm the main phonon-energy window and uncover an additional reduced window that separates momentum-space regions with an excess and deficit in the electronic population.
We also identify a separate bottleneck in the phonon thermalization, which is rooted in the momentum-dependent coupling to the particle-hole continuum.
The high momentum and frequency resolution of our spectra makes the underlying phonon--charge correspondence directly visible.
For acoustic phonons, the phonon-energy window acquires pronounced momentum dependence dictated by simultaneous energy and momentum conservation.
The reduced window becomes asymmetric, and directional scattering between Brillouin-zone regions produces a particularly persistent bottleneck for low-momentum phonon modes.
Our results establish diagrammatic QTT-NEGF simulations as a scalable and controlled framework for quantitative nonequilibrium electron-phonon dynamics, which overcomes previous lattice-size and propagation-time limitations, and provides accurate reference data for time-resolved spectroscopies.
\end{abstract}

\maketitle
	
\section{Introduction}

In pump-probe experiments, phonons are crucial degrees of freedom which strongly influence the dynamics of photo-excited solids from subpicosecond to picosecond timescales.
Phenomenological descriptions of the transient evolution of an electron-phonon coupled system often assume separate temperatures of the electron and phonon subsystems, the so-called two-temperature model \cite{Kaganov1957,Ginzburg1955,Lifshits1960,Anisimov1974,Allen1987} or its extensions, the multi-phonon-temperature models or nonthermal lattice models \cite{Waldecker2016,Maldonado2017,Lu2018,Caruso2020,Novko2020,Tajik2023}.
However, these simplified treatments fail to capture phenomena associated with general nonthermal electron and phonon distributions, such as transient population inversions \cite{Werner2016} or relaxation bottlenecks \cite{Sentef2013,Kemper2014,Murakami2015,Rameau2016,Abdurazakov2018,Kemper2018}. 
While more detailed insights can be obtained from microscopic model studies \cite{Werner2013,Murakami2015,Picano2023}, and proper ab~initio electron-phonon formalisms have recently been developed \cite{Stefanucci2023,Stefanucci2024}, our understanding of the nonequilibrium dynamics even in the most basic electron-phonon systems remains incomplete.

This knowledge gap originates from the difficulty of accurately simulating the many-body dynamics of electron-phonon coupled systems.
Wave-function-based techniques, such as exact diagonalization or the density-matrix renormalization group, face severe challenges because of the (in principle) infinite size of the phonon Hilbert space \cite{Jeckelmann2007}. Dealing with this in practice requires careful truncations \cite{Caron1996,Caron1997}, pseudosite formulations \cite{Jeckelmann1998,Jeckelmann1999,Ge2024}, or special basis optimizations \cite{Zhang1998,Zhang1999,Bonca1999,Barford2002,Wong2008,Guo2012,Golez2012,Brockt2015,Dorfner2015,Schroder2016,Brockt2017,Stolpp2020}.
This greatly restricts the maximum propagation times, and in particular, the lattice size and dimensionality (most previous studies considered one dimension).
While diagrammatic techniques bypass this issue by replacing wave functions by Green's functions, in nonequilibrium, the cost of storing the two-time electron and phonon Green's functions, as well as the corresponding self-energies, quickly becomes prohibitive with increasing propagation time.
The computational effort also rapidly grows due to the non-Markovian character of the equations of motion.
The above issues become particularly challenging when accounting for the phonon's dispersion, which requires storing a large two-time matrix at each momentum point for each propagator and self-energy.

Previous nonequilibrium Green's function (NEGF) studies with full two-time dependence therefore focused on nondispersive phonons, typically within dynamical mean-field theory (DMFT)~\cite{Aoki2014} or related local self-energy approximations~\cite{Sentef2013,Werner2013,Kemper2014,Murakami2015,Rameau2016,Abdurazakov2018,Kemper2018}.
These works revealed nontrivial effects, such as phonon echos \cite{Werner2013}, a crossover from electron- to phonon-dominated relaxation \cite{Murakami2015}, and bottlenecks in the electron relaxation \cite{Sentef2013}.
Still, some of these studies \cite{Sentef2013,Kemper2014,Rameau2016,Kemper2018} assumed the phonon subsystem to have infinite heat capacity, so that the system relaxed back to the cold initial state after absorbing the excitation energy.
In reality, the bidirectional energy transfer between electrons and phonons controls the relaxation~\cite{Konstantinova2018}, and a consistent treatment of this mutual feedback requires renormalized phonons.

Dispersive phonons, whose momentum-dependent self-energies make a full NEGF treatment very expensive, have instead been studied with simpler methods.
These include the master-equation approach~\cite{Schueler2017}, first-principles time-local equations of motion~\cite{Mocatti2026}, time-dependent Boltzmann equations~\cite{Rethfeld2002,Mueller2013,Baranov2014,Kim2011,Jhalani2017,Ono2020,Tong2021,Caruso2021,Seiler2021,Caruso2022,Held2025}, kinetic equations \cite{Vasko2008}, the generalized Kadanoff-Baym ansatz with a completed collision approximation~\cite{Marini2013}, related time-local approximations~\cite{Sangalli2015}, and real-time time-dependent density functional theory~\cite{Miyamoto2006,Habenicht2008,Guan2022}.
These studies revealed the role of phonon dissipation in topological phase transitions~\cite{Schueler2017} and described nonequilibrium electron thermalization, delayed cooling, and energy transfer in metals beyond the two-temperature model~\cite{Rethfeld2002,Mueller2013,Baranov2014}. They also considered ultrafast photoluminescence in metals \cite{Ono2020}, photoconductivity in graphene \cite{Vasko2008}, hot-carrier relaxation and population inversions in graphene and wide-gap semiconductors~\cite{Kim2011,Jhalani2017,Tong2021}, and highly anisotropic nonthermal phonon populations \cite{Caruso2021,Seiler2021} which can even collapse the electron-phonon energy transfer rate \cite{Held2025}.
The time-local approaches \cite{Marini2013,Sangalli2015} studied the importance of different scattering channels in semiconductors, while real-time time-dependent density functional theory provided ab~initio simulations of nonequilibrium electron-phonon dynamics in nanotubes, quantum dots, and monolayers~\cite{Miyamoto2006,Habenicht2008,Guan2022}.

While these studies revealed important nonequilibrium effects in systems with dispersive phonons, the methods employed do not capture the full two-time quantum dynamics and the self-consistent feedback between electrons and phonons that a conserving diagrammatic approach provides.
The effect of phonon dispersion on the nonequilibrium evolution has therefore not yet been systematically explored within a fully quantum, self-consistent NEGF framework.

In this paper, we fill this gap by comparing the relaxation of electron-phonon systems with optical and acoustic phonon dispersions using the renormalized Migdal approximation on lattices of up to $256\times 256$ sites and up to $t_\mathrm{max}=200$ inverse hopping times.
This is a substantial advance over previous full two-time NEGF simulations of two-dimensional electron-phonon systems, which were constrained to optical phonons within a local self-energy approximation, momentum grids of about $100\times100$ points, and shorter times \cite{Kemper2013HHG,Abdurazakov2018}.
Our calculations are made possible by the recently developed quantics-tensor-train (QTT) NEGF approach \cite{Sroda2024,Inayoshi2025,Sroda2025}, which dramatically reduces the memory demand for storing momentum-resolved two-time Green's functions by employing a tensor-train form.
This allows us to retain the full two-time Green's functions and use them directly in the collision integrals.
By resolving the momentum-dependent electron, hole, and phonon densities as well as the self-energies over long propagation times ($t_\mathrm{max}=200$), we reveal a rich hierarchy of relaxation bottlenecks that extends and refines the phonon-window effect discussed in earlier works~\cite{Sentef2013,Kemper2014,Murakami2015,Rameau2016,Abdurazakov2018,Kemper2018}.
For optical phonons, we confirm the phonon-window bottleneck, a long-lived nonthermal electron population within the phonon energy range, and we uncover an additional reduced window that separates regions of enhanced and suppressed nonthermal electron population.
At early times, we furthermore resolve transient higher-order windows where the fermion relaxation is additionally delayed.
We also identify a separate bottleneck in the phonon thermalization, rooted in the momentum-dependent coupling to the particle-hole continuum, which dictates which phonon modes thermalize quickly and which remain frozen.
The high momentum and frequency resolution of our spectra makes the underlying correspondence between the phonon and charge response, previously established in equilibrium~\cite{Engelsberg1963,Engelsberg1964,Sykora2006,Assaad2008,Hohenadler2011,Weber2015,Walther2026}, directly visible.
For acoustic phonons, the phonon-energy window persists, even though acoustic modes can in principle absorb arbitrarily small energies from the fermions. 
In this case, the window acquires a pronounced momentum dependence dictated by simultaneous energy and momentum conservation.
The reduced window becomes asymmetric, and directional scattering between different regions of the Brillouin zone, together with the vanishing of the electron-phonon coupling for low momenta, produces a particularly persistent bottleneck for low-momentum phonon modes.
This inefficient low-momentum emission causes the long-lived nonthermal fermion population in the acoustic case, and directly links the phonon relaxation bottleneck to the fermion relaxation bottleneck.

This paper is organized as follows.
Section~\ref{sec:model} introduces the model, the phonon dispersions, and the QTT-NEGF methodology.
In Sec.~\ref{results}, we present the simulation results: first the fermion (Sec.~\ref{optical_fermion_relaxation}) and phonon (Sec.~\ref{optical_ph_relaxation}) relaxation for optical phonons, followed by the fermion (Sec.~\ref{acoustic_fermion_relaxation}) and phonon (Sec.~\ref{acoustic_ph_relaxation}) relaxation for acoustic phonons.
Section~\ref{conclusions} summarizes our findings and discusses their implications for time-resolved spectroscopies and future extensions of the method.

\section{Model and method}\label{sec:model}

\subsection{Model and diagrammatic NEGF approach}

\begin{figure}[t]
	\centering
	\includegraphics[width=\columnwidth]{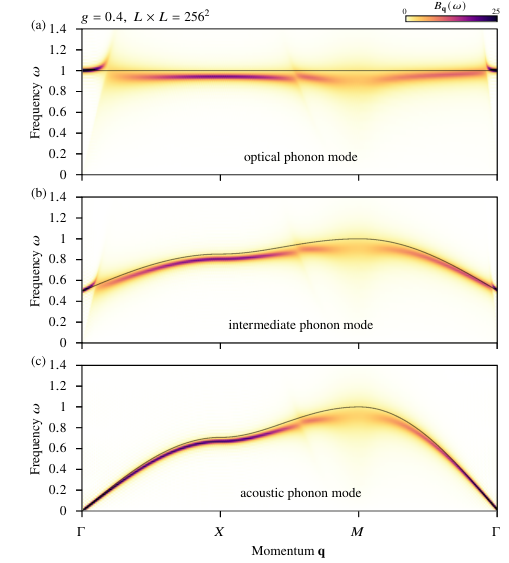}
	\caption{Bare and renormalized equilibrium phonon dispersions, $\Omega_\mathbf{q}$ (lines) and $B_\mathbf{q}(\omega)=(-1/\pi)\,\mathrm{Im}\,D^R_\mathbf{q}(\omega)$ (color maps), for (a) optical, (b) intermediate ($x=1/2$), and (c) acoustic modes.
	The e-ph coupling is $g=0.4$, the lattice size is $L\times L=256^2$, and the inverse temperatures are (a) $\beta=5.2598$, (b) $\beta=5.8382$, and (c) $\beta=6.7671$, corresponding to the thermalized states after the quench.
	The equilibrium calculations are performed by starting from the Matsubara solution and propagating the Kadanoff-Baym equations on the real-time axis up to $t_\text{max}=200$.
	}
	\label{fig1}
\end{figure}

We consider a half-filled electron-phonon (e-ph) system on a two-dimensional $L \times L$ square lattice, described by the Fröhlich Hamiltonian
\begin{equation}
    \begin{aligned}
        H(t) =&~H_\mathrm{e} + H_\mathrm{ph} + H_\text{e-ph}(t)\\[.5em]
        =& \sum_{\mathbf{k}\sigma} \epsilon_{\mathbf{k}} c^\dagger_{\mathbf{k}\sigma} c_{\mathbf{k}\sigma} + \sum_{\mathbf{q}} \Omega_\mathbf{q} b^\dagger_{\mathbf{q}} b_{\mathbf{q}} \\
		&+ \frac{1}{\sqrt{N_\mathbf{k}}}\sum_{\mathbf{kq}\sigma} g_\mathbf{q}(t) c^\dagger_{\mathbf{k+q}\,\sigma}c_\mathbf{k\sigma} (b_{-\mathbf{q}}^\dagger + b_\mathbf{q}).
    \end{aligned}\label{eq:holstein}
\end{equation}
Here, $c^\dagger_{\mathbf{k}\sigma}$ creates an electron with momentum $\mathbf{k}=(k_x,k_y)$ and spin $\sigma$, whereas $b^\dagger_{\mathbf{q}}$ creates a phonon with momentum $\mathbf{q}=(q_x,q_y)$.
The electron dispersion is $\epsilon_{\mathbf{k}}=-2t_\mathrm{h}(\cos{k_x}+\cos{k_y})$ and $N_\mathbf{k}=L^2$ is the number of $\mathbf{k}$ points.
Our numerical approach captures arbitrary phonon dispersions; we illustrate this flexibility in Fig.~\ref{fig1} with the equilibrium renormalized phonon spectra for three representative choices of the bare dispersion
\begin{equation}
	\Omega_\mathbf{q} =
	\begin{cases}
		\Omega_0, & \text{(optical)},\\[.5em]
		\frac{\Omega_0}{\sqrt{2}}\sqrt{\sin^2 \frac{q_x}{2} + \sin^2 \frac{q_y}{2}} & \text{(acoustic)},\\[.5em]
		x \Omega_0 + (1-x) \frac{\Omega_0}{\sqrt{2}}  \sqrt{\sin^2 \frac{q_x}{2} + \sin^2 \frac{q_y}{2}} & \text{(intermediate)},
	\end{cases}
\end{equation}
These span the limiting cases of purely optical (dispersionless) and acoustic (dispersive) phonons, together with an intermediate case at $x=1/2$.
In what follows we focus on the extremal optical and acoustic cases, since the intermediate case follows from these.
Namely, we find that the fermion relaxation is qualitatively closer to the optical case, albeit with a weak asymmetry in the population excess/deficit pattern, whereas the phonon relaxation exhibits features of both limiting dispersions: slow relaxation of essentially unrenormalized modes, as in the optical case, and slow emission of low-$\mathbf{q}$ phonons, as in the acoustic case.
The e-ph coupling is taken as
\begin{equation}
	g_\mathbf{q}(t) = \sqrt{\Omega_\mathbf{q}}\, g(t),
\end{equation}
which for acoustic phonons correctly behaves as $g_\mathbf{q} \xrightarrow{q \to 0} \sqrt{q}$.
The system is excited by a quench of the coupling $g(t < 0) = 0 \to g(t \geq 0) = g$.
For the slow bottleneck dynamics studied here, the coupling quench is expected to lead to the same conclusions as a broadband laser excitation, while being more convenient numerically.
The energy unit is the hopping amplitude $t_\mathrm{h}=1$ (corresponding to the bandwidth $W=8$), the time unit is $1/t_\mathrm{h}$ ($\hbar=1$), and we consider paramagnetic states.
We further fix $\Omega_0=1$, which is chosen large to clearly distinguish different phonon-energy windows \footnote{The corresponding $\Omega_0/W$ ratio is in the realistic range for intramolecular vibrations in fullerides and $\kappa$-organic compounds \cite{Nomura2016, Buzzi2020}.}. 
We present results for $L\times L = 128^2$ and $256^2$.

To study the dynamics, we calculate the interacting nonequilibrium electron and phonon (displacement) Green's functions, 
\begin{equation}
	G_\mathbf{k}(t,t')=-\im\langle \mathcal{T}_\mathcal{C} c_\mathbf{k}(t)c_\mathbf{k}^\dagger(t') \rangle\label{eq:G}
\end{equation}
and
\begin{equation}
D_\mathbf{q}(t,t')=-\im\langle \mathcal{T}_\mathcal{C} X_\mathbf{q}(t)X_\mathbf{q}^\dagger(t') \rangle,\label{eq:D}
\end{equation}
where $X_\mathbf{q} = b_{-\mathbf{q}}^\dagger + b_\mathbf{q}$ is the phonon displacement operator and $\mathcal{T}_\mathcal{C}$ is the time-ordering operator on the Kadanoff-Baym contour, $\mathcal{C} \equiv \includegraphics[height=.7em]{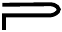}$, which runs along $0 \to t_\text{max} \to 0 \to -\im\beta$ with the maximum time $t_\text{max}$ and inverse temperature $\beta$ \cite{Aoki2014}.
To analyze time-dependent spectra, we use a partial (Wigner) transform of generic two-time functions $Y(t_1,t_2)$,
\begin{equation}
	Y(\omega, t) = \int dt_\mathrm{rel}\, e^{\im \omega t_\mathrm{rel}}\, Y\!\left(t + \tfrac{1}{2}t_\mathrm{rel},\, t - \tfrac{1}{2}t_\mathrm{rel}\right),\label{eq:wigner}
\end{equation}
where $t=(t_1+t_2)/2$ and $t_\mathrm{rel}=t_1-t_2$ are the average and relative times, respectively, and in practice a Gaussian window with standard deviation $\delta = t_\text{max}/2$ is applied to the $t_\mathrm{rel}$ integral.

The e-ph interactions are treated within the renormalized Migdal approximation \cite{Migdal1958,Murakami2015,Abdurazakov2018}. The electron self-energy $\Sigma$ in real space reads
\begin{equation}
	\begin{aligned}
		\Sigma_{ij}(t,t') &= \vcenter{\hbox{
			\begin{tikzpicture}[baseline={(current bounding box.center)}]
			\begin{feynman}
				\vertex (a);
				\vertex[right=1.5cm of a] (b);
		
				\diagram*{ (a) -- [fermion, thick] (b), };
				\draw[boson, thick] (a) .. controls +(0,0.8) and +(0,0.8) .. (b);
		
			\end{feynman}
			\end{tikzpicture}
		}}
		= \im G_{ij}(t,t') V_{ij}(t,t') ,
	\end{aligned}
\end{equation}
where $V$ denotes the phonon-mediated electron-electron interaction, given in momentum space by $V_\mathbf{q}(t,t') = g_\mathbf{q}(t) D_\mathbf{q}(t,t') g_\mathbf{q}(t')$. The real-space functions are obtained by the Fourier transform $f_{ij}(t,t')=(1/N_\mathbf{k}) \sum_\mathbf{k} e^{-\im\mathbf{k}(\mathbf{r}_i-\mathbf{r}_j)} f_\mathbf{k}(t,t')$. The phonon self-energy $\Pi$ is defined as
\begin{equation}
		\Pi_{\mathbf{q}}(t,t') = g_\mathbf{q}(t) P_{\mathbf{q}}(t,t') g_\mathbf{q}(t'),\label{eq:pi}
\end{equation}
where $P$ denotes the polarization bubble, given in real space by
\begin{equation}
	P_{ij}(t,t') = \vcenter{\hbox{
		\begin{tikzpicture}[baseline={(current bounding box.center)}]
		\begin{feynman}
			\vertex (a);
			\vertex[right=1.5cm of a] (b);
	
			\draw[fermion, thick] (a) .. controls +(0,.55) and +(0,.55) .. (b);
			\draw[anti fermion, thick] (a) .. controls +(0,-.55) and +(0,-.55) .. (b);
		\end{feynman}
		\end{tikzpicture}
	}} =- 2\im G_{ij}(t,t') G_{ji}(t',t).\label{eq:P}
\end{equation}
The self-energies enter the fermionic and phononic Dyson equations,
\begin{align}
	G_\mathbf{k}(t,t') &= G_{0\mathbf{k}}(t,t') + [G_{0\mathbf{k}} * \Sigma_\mathbf{k} * G_\mathbf{k}](t,t')\label{eq:dysonG},\\
	D_\mathbf{q}(t,t') &= D_{0\mathbf{q}}(t,t') + [D_{0\mathbf{q}} * \Pi_\mathbf{q} * D_\mathbf{q}](t,t'),\label{eq:dysonD}
\end{align}
whose solution yields the Green's functions \eqref{eq:G} and \eqref{eq:D}. In the actual implementation, we solve the above equations component-wise, i.e., as four coupled Kadanoff-Baym equations (KBE) for the Matsubara ($M$), retarded ($R$), left-mixing ($\rceil$) and lesser ($<$) components \cite{Aoki2014}. $[f*g](t,t') = \int_\mathcal{C} d\bar{t} f(t,\bar{t}) g(\bar{t},t')$ is a contour convolution and the noninteracting Green's functions are given by \cite{Freericks2005}
\begin{align}
	G_{0\mathbf{k}}(t,t') = &-\im[\theta_\mathcal{C}(t,t')-f_\beta(\epsilon_\mathbf{k})]e^{-\im\epsilon_\mathbf{k} (t-t')},\\
	D_{0\mathbf{q}}(t,t') = &-\im[\theta_\mathcal{C}(t,t')+b_\beta(\Omega_\mathbf{q})]e^{-\im\Omega_\mathbf{q}(t-t')}\\
	&+ \im[\theta_\mathcal{C}(t,t')+b_\beta(-\Omega_\mathbf{q})]e^{\im\Omega_\mathbf{q}(t-t')},\nonumber
\end{align}
where $f_\beta(\omega)$ [$b_\beta(\omega)$] represents the Fermi-Dirac [Bose-Einstein] distribution for the initial inverse temperature $\beta$ and $\theta_\mathcal{C}(t,t')$ represents the contour step function \cite{Aoki2014}.

Equations \eqref{eq:dysonG} and \eqref{eq:dysonD} are solved self-consistently, which results in a conserving approximation and captures the mutual feedback between the electron and phonon subsystems.
This feedback leads to a non-trivial renormalization of the phonon dispersions, as shown in Fig.~\ref{fig1} for some representative equilibrium systems and further discussed in Sec.~\ref{optical_ph_relaxation}.
This, together with the possibly nontrivial bare phonon dispersion, necessitates a fully momentum-dependent treatment of both the electron and phonon Green's functions.
We achieve this on a fine momentum grid (large lattice) by employing the recently developed QTT-NEGF method \cite{Sroda2024,Inayoshi2025,Sroda2025}.

\subsection{QTT-NEGF}

The idea behind QTT-NEGF is to mitigate the prohibitive memory demand of storing the momentum-dependent two-time correlators by compressing them into tensor trains \cite{MSSTA}.
Then, the correlators are manipulated exclusively in this format with the help of the established tensor-network toolbox.
Crucially, the method retains the full two-time Green's functions and uses them directly in the collision integrals, rather than reconstructing the off-diagonal components.
For the compression, the so-called quantics representation is invoked \cite{MSSTA}.
It ``tensorizes'' the standard matrix representation of $G_\mathbf{k}(t,t')$ (resulting from the discretization of the contour $\mathcal{C}$) by enumerating the time grid points $t$, $t'$ with binary numbers $(t_1,\ldots,t_R)_2, (t'_1,\ldots,t'_R)_2 \in \{{0,1,\ldots,2^R-1}\}$.
Each binary digit $t_n, t'_n \in \{0,1\}$ becomes an external leg in the tensor train $\begin{tikzpicture}
	\begin{pgfonlayer}{nodelayer}
		\node [style=qtt] (0) at (-2, 0.125) {};
		\node [style=qtt] (1) at (-1, 0.125) {};
		\node [style=none] (4) at (0, 0.125) {$\ldots$};
		\node [style=qtt] (5) at (1, 0.125) {};
		\node [style=qtt] (6) at (2, 0.125) {};
		\node [style=none] (7) at (-2, -0.4) {};
		\node [style=none] (8) at (-1, -0.4) {};
		\node [style=none] (12) at (1, -0.4) {};
		\node [style=none] (13) at (2, -0.4) {};
		\node [style=none] (14) at (-0.5, 0.125) {};
		\node [style=none] (15) at (0.5, 0.125) {};
	\end{pgfonlayer}
	\begin{pgfonlayer}{edgelayer}
		\draw [style=normal] (0) to (7.center);
		\draw [style=normal] (1) to (8.center);
		\draw [style=normal] (5) to (12.center);
		\draw [style=normal] (6) to (13.center);
		\draw [style=normal] (6) to (5);
		\draw [style=normal] (1) to (0);
		\draw [style=normal] (5) to (15.center);
		\draw [style=normal] (1) to (14.center);
	\end{pgfonlayer}
\end{tikzpicture}
$, which is 
a factorized representation of the original matrix.
Crucially, while the compression is lossy, its faithfulness is fully controllable by an adjustable cutoff tolerance $\tau_\mathrm{cutoff}$ that limits the allowed error in terms of the Frobenius norm, $|f-\tilde{f}|_\mathrm{F}^2/|f|_\mathrm{F}^2 < \tau_\mathrm{cutoff}$ ($f$, $\tilde{f}$ being the original and compressed functions).
This cutoff fixes the dimensions $D$ of the extra ``bond'' indices that are multiplied over in the factorization. 
Importantly, $D$ is found to be moderate, $\mathcal{O}(100)$, for a large class of functions, including many physical correlation functions \cite{MSSTA}, leading to huge memory savings, as a QTT stores only $\mathcal{O}(D^2\times\mathrm{length})$ elements.

Due to the binary encoding, the magnitude of $D$ reflects the level of correlation, or ``entanglement'' \cite{Rohshap2025a,Rohshap2025b}, between different time scales on which the compressed function varies.
This intuitively explains the large compressibility through the concept of scale separation \cite{MSSTA,Ritter2024,Rohshap2025a,Rohshap2025b}.
The exponential character of the binary encoding provides one more advantage: one can work with very fine grids $dt\sim 10^{-7}$, since the bond dimensions on the fine-scale tensors become trivial, $D=1$, when all the features of the function are resolved.
The possibility to use such fine grids facilitates the implementation, as one can perform accurate integrations with low-order quadrature rules.

Once all one-time and two-time functions appearing in the NEGF formalism are cast into the QTT format, one can manipulate them without ever decompressing them.
The preparation of noninteracting Green's functions is done directly [$\mathcal{O}(1)$] or via tensor cross interpolation \cite{Oseledets2011,Dolgov2020,Fernandez2022,Ritter2024,Fernandez2024} [$\mathcal{O}(D^3)$]. 
Matrix multiplications (discretized convolutions) and element-wise multiplications (needed to calculate diagrams) correspond to tensor network contractions [$\mathcal{O}(D^4)$].
Summations are simply tensor-train sums [$\mathcal{O}(D^3)$].
Finally, the solution of Eqs.~\eqref{eq:dysonG}-\eqref{eq:dysonD} can be done by viewing them as a linear problem and performing a sweeping algorithm that sequentially updates individual tensors of the train by solving small local linear problems [$\mathcal{O}(D^4)$].

For further details of the QTT-NEGF implementation used here, see Refs.\ \cite{Sroda2024,Inayoshi2025,Sroda2025}, whereas for a more generic discussion of the applicability of QTTs to many-body correlation functions, including the discussion of standard linear-algebra operations, see Ref.~\cite{MSSTA}. The implementation is written in Julia \cite{bezanson2017julia} and is based on the ITensor library \cite{itensor} and libraries developed by the tensor4all group \cite{Fernandez2024,tensor4all}.

\subsection{Implementation details}

To take advantage of causality, the solution of Eqs.~\eqref{eq:dysonG}-\eqref{eq:dysonD} is block-time-stepped \cite{Inayoshi2025,Sroda2025}.
One solves the equations one by one on moderately large block time steps $dt_\mathrm{step}\sim \mathcal{O}(10)$ that still encompass a huge number of underlying grid points with fine spacing $dt\sim 10^{-7}$.
The self-consistency on a block time step is reached when $\mathrm{max}_\mathbf{k} \sum_{c=R,\rceil,\gtrless} {|G^c_{\mathbf{k},\mathrm{new}}-G_{\mathbf{k},\mathrm{old}}^{c}|_\mathrm{F}}/{|G_{\mathbf{k},\mathrm{old}}^c|_\mathrm{F}} < \epsilon_\mathrm{conv}$ (the Frobenius norms are calculated on the time domain including the converged past data and the current block time step).
Since there are only few relevant timescales controlling the evolution within a given block time step, this propagation scheme helps both to stabilize the equations and to control the computational cost by reducing iteration counts and the bond dimensions.

To efficiently simulate large system sizes, we step up through progressively larger lattices.
First, at $L\times L = 64^2$, we proceed as described above, propagating the KBE with block time stepping to the desired $t_\mathrm{max}$ and converging each step below $\epsilon_\mathrm{conv}=1 \times 10^{-3}$.
Next, we Fourier interpolate the self-energy $\Sigma_\mathbf{k}$ to $L\times L = 128^2$, use it to obtain the Green's functions $G_\mathbf{k}$ and $D_\mathbf{q}$ from Eqs.~\eqref{eq:dysonG} and \eqref{eq:dysonD} \cite{Moritz2010,Nowadnick2015}, and apply global contour updates \cite{Sroda2024,Gasperlin2025}, reaching self-consistency at $\epsilon_\mathrm{conv} \approx 5 \times 10^{-3}$.
Typically, 2-3 iterations are enough to converge, since the interpolated function is very close to the true solution (the error in the first iteration is only $\sim 10^{-1}$).
Finally, for $L\times L = 256^2$, we proceed analogously but omit the global contour updates, which become very costly at this lattice size, and calculate the observables based on the interpolated self-energy.
We typically use a tolerance $\tau_\mathrm{cutoff}=10^{-11}$--$10^{-10}$ and maximum bond dimension $D_\mathrm{max}=60$--$80$ for the $L\times L=64^2$ and $128^2$ calculations.
For $L\times L = 256^2$, we instead use $\tau_\mathrm{cutoff} = 10^{-9}$--$10^{-8}$ and $D_\mathrm{max} = 50$ so that the runs remain efficient.

Each simulation was performed on a single dual-socket computational node with two 64-core AMD EPYC 7742 2.25 GHz processors and 768 GB of random access memory.
This modest hardware requirement demonstrates the effectiveness of the QTT compression in the context of NEGF simulations.

\section{Results}\label{results}

\begin{video*}[p]
	\centering
	\includegraphics[width=\textwidth]{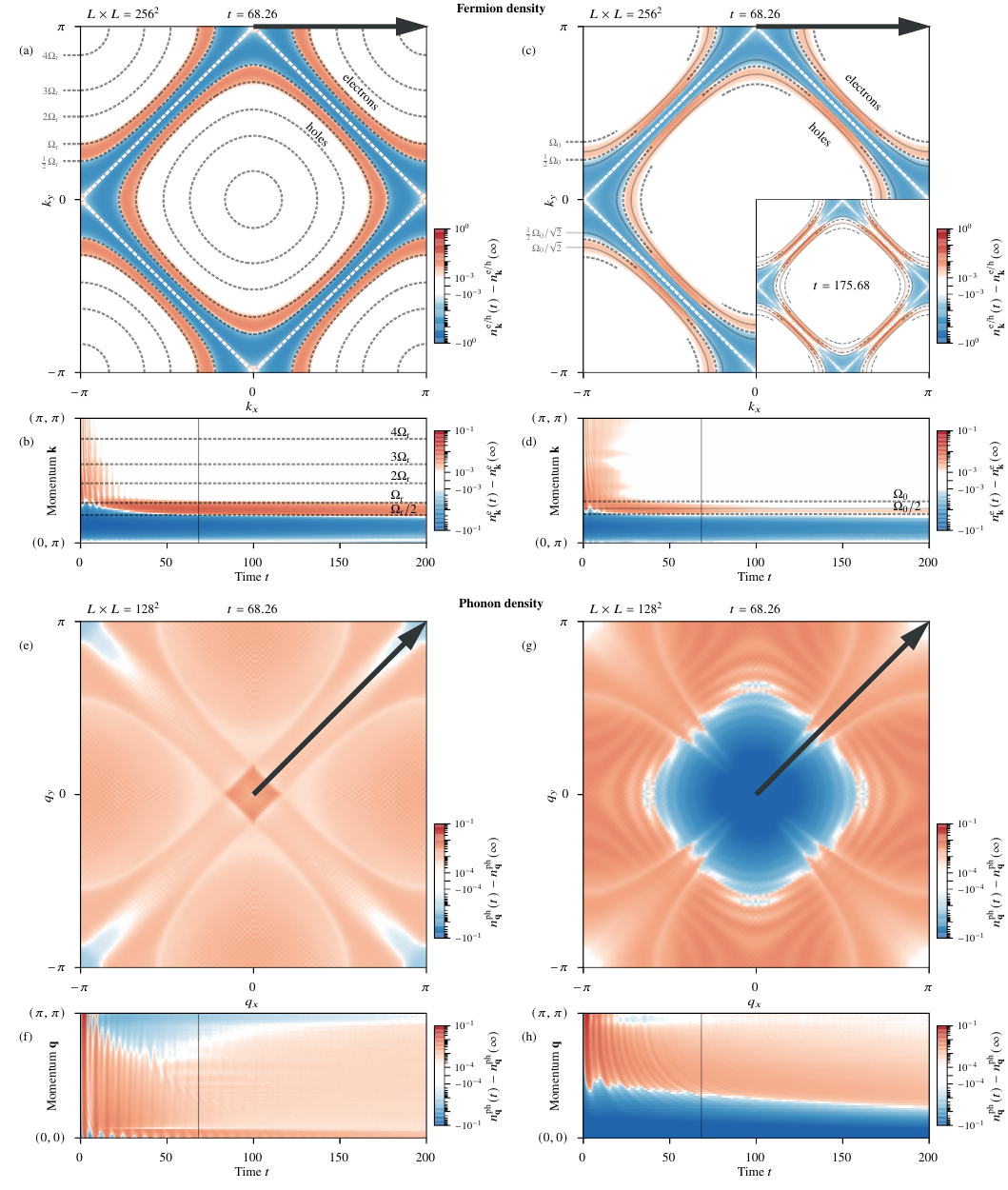}
	\setfloatlink{\SupplementaryMultimediaURL}
	\caption{\label{vid1}%
		Momentum-resolved deviation $n^{\mathrm{e,h,ph}}_\mathbf{k}(t)-n^{\mathrm{e,h,ph}}_\mathbf{k}(\infty)$ of the fermion and phonon populations from their thermalized values after a $g=0{\to}0.4$ quench ($\beta=10$), for optical (left) and acoustic (right) phonons; red (blue) marks an excess (deficit).
		Panels (a),(c) show the fermion-density deviation over the Brillouin zone ($L\times L=256^2$) at $t=68.26$: $n^\mathrm{e}_\mathbf{k}$ outside the Fermi surface (white dashed), $n^\mathrm{h}_\mathbf{k}$ inside it. The inset in (c) shows the $t=175.68$ result.
		Panels (b),(d) show the same data as a function of time and along the $(0,\pi){\to}(\pi,\pi)$ path [arrows in (a),(c)].
		Panels (e),(g) show the phonon-density deviation over the Brillouin zone ($L\times L=128^2$, due to the accuracy required to resolve deviations down to $10^{-4}$) at $t=68.26$.
		Panels (f),(h) show the same data as a function of time and along the $(0,0){\to}(\pi,\pi)$ path [arrows in (e),(g)].
		Dashed lines in (a)-(d) mark the constant band energies $\epsilon_\mathbf{k}=(1/2)\Omega_\mathrm{r}$ and $n\Omega_\mathrm{r}$ ($n \in \mathbb{Z}$; optical) or $\epsilon_\mathbf{k}=(1/2)\Omega_0$ and $\Omega_0$ (acoustic). Solid lines in (c) mark $\epsilon_\mathbf{k}=(1/2)\Omega_0/\sqrt{2}$ and $\Omega_0/\sqrt{2}$ [also shown in (d)].
		Vertical solid lines in (b),(d),(f),(h) mark the snapshot time $t=68.26$.
		Full time-resolved data are available as supplementary multimedia \cite{Vid1Supp}.
		}
\end{video*}

We study the thermalization after a sudden quench of the e-ph coupling from $g=0$ to $g=0.4$, starting from an initial inverse temperature $\beta=10$.
The resulting momentum-resolved relaxation is summarized in Video~\ref{vid1} (use the link in Ref.~\cite{Vid1Supp} to play the video), which shows the time-dependent deviation of the fermion (electron or hole) and phonon densities from their thermalized ($t\to\infty$) values, $n^{\mathrm{e,h,ph}}_\mathbf{k}(t)-n^{\mathrm{e,h,ph}}_\mathbf{k}(\infty)$.
The electron and hole densities are defined as $n^\mathrm{e}_\mathbf{k}(t)=-\im G_\mathbf{k}^<(t,t)$ and $n^\mathrm{h}_\mathbf{k}(t)=\im G_\mathbf{k}^>(t,t)$, respectively, and the phonon density as
\begin{align}
	n^\mathrm{ph}_\mathbf{q}(t) &= \langle b_\mathbf{q}^\dagger(t) b_\mathbf{q}(t) \rangle \nonumber\\
	&= \frac{\im}{4}\left( D_\mathbf{q}^<(t,t') + \frac{1}{\Omega^2_\mathbf{q}}\left[\partial_t \partial_{t'} D_\mathbf{q}(t,t')\right]^< \right)\Bigg\vert_{t'=t} - \frac{1}{2}.\label{eq:nph}
\end{align}
The derivatives of the phonon displacement Green's function $D_\mathbf{q}(t,t')$ are evaluated as described in the Appendix of Ref.~\cite{Murakami2015}.
To calculate differences in observables relative to the thermalized state, we identify the temperature of that state by matching its energy to the post-quench total energy. Since the Migdal approximation is conserving, the total energy remains constant during the time evolution, which justifies this identification.
Panels (a) and (c) display the density deviations across the Brillouin zone (BZ) at a fixed time, for the optical and acoustic phonons, respectively, while panels (b) and (d) show the data along the path $(0,\pi)\to(\pi,\pi)$ as a function of time.
Outside the Fermi surface we plot the electron-density deviation, and inside it the hole-density deviation; red marks an excess and blue a deficit relative to the thermalized state.
Panels (e)-(h) show the phonon data in an analogous manner.

At $t=0$ the system is in a cold initial state, 
with a fairly sharp Fermi surface and essentially no excited phonons.
Because the final thermalized state is hotter and its distributions are correspondingly broader, the deviations $n^{\mathrm{e,h,ph}}_\mathbf{k}(t)-n^{\mathrm{e,h,ph}}_\mathbf{k}(\infty)$ are initially negative (blue) throughout the BZ for electrons, holes, and phonons alike.
That is, there is a deficit of electrons above the Fermi surface, a deficit of holes below it, and a deficit of phonons throughout the BZ.

\subsection{Results for the model with optical phonons}

\subsubsection{Fermion relaxation (optical phonons)}\label{optical_fermion_relaxation}

Let us first focus on the fermion thermalization in the model with optical phonon modes [Video~\ref{vid1}(a) and (b)].
The quench acts as a broadband perturbation and almost instantaneously generates a nonthermal electron population across the entire BZ, together with a small phonon population.
The high-energy electrons and holes then rapidly relax toward the Fermi level by emitting phonons.
However, their relaxation slows down dramatically at momenta which are clearly separated from the Fermi surface (located at $\epsilon_\mathbf{k}=0$). Consequently, a long-lived nonthermal fermion population builds up, marked by a nonzero $n^{\mathrm{e,h}}_\mathbf{k}(t)-n^{\mathrm{e,h}}_\mathbf{k}(\infty)$.
This long-lived population is a manifestation of the electron relaxation bottleneck identified in Ref.~\cite{Sentef2013}.
The momenta with a persisting nonthermal fermion occupation map onto the phonon energy window $\mathcal{W}=[-\Omega_\mathrm{r},\Omega_\mathrm{r}]$ (dashed contours in Video~\ref{vid1}), where $\Omega_\mathrm{r}$ is the renormalized phonon frequency, which we identify with the peak of the momentum-averaged phonon spectral function $\sum_\mathbf{q} B_\mathbf{q}(\omega,t\to\infty)=\sum_\mathbf{q} (-1/\pi)\,\mathrm{Im}\,D^R_\mathbf{q}(\omega,t\to\infty)$.

\begin{figure*}[tb]
	\centering
	\includegraphics[width=\textwidth]{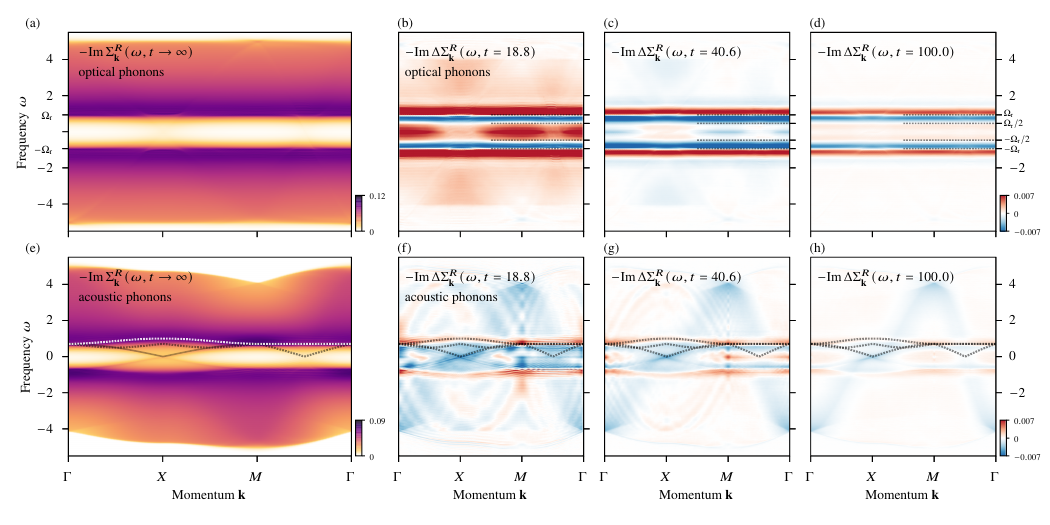}
	\caption{Imaginary part of the retarded electron self-energy $-\mathrm{Im}\,\Sigma^R_\mathbf{k}$ for (a)-(d) optical and (e)-(h) acoustic phonons.
	Panels (a),(e) show the equilibrium self-energy $-\mathrm{Im}\,\Sigma^R_\mathbf{k}(\omega,t\to\infty)$ for the thermalized post-quench state with $g=0.4$ and $\beta=5.2598$ (optical) or $\beta=6.7671$ (acoustic).
	Panels (b)-(d) and (f)-(h) show the time-dependent deviation from this equilibrium result, $-\mathrm{Im}\,\Delta\Sigma^R_\mathbf{k}(\omega,t)$, at $t=18.8$,$40.6$,$100.0$ after quenches $g=0{\to}0.4$ with initial $\beta=10$.
	In (a), the $y$-axis ticks mark $\pm\Omega_\mathrm{r}$, with $\Omega_\mathrm{r}$ the renormalized phonon frequency, while in (b)-(d), the right-hand $y$-axis ticks mark $\pm\Omega_\mathrm{r}$ and $\pm\Omega_\mathrm{r}/2$.
	In (e), the lines indicate different scattering channel contributions: $\omega_{\mathbf{k}\to X}$ (black solid), $\omega_{\mathbf{k}\to(\frac{\pi}{2},\frac{\pi}{2})}$ (black dashed), and $\omega_{\mathbf{k}\to X'}$ (white dashed); they are overlaid also on (f)-(h).
	Here, $L\times L=256^2$ throughout.
	}
	\label{fig2}
\end{figure*}

The existence of this phonon-energy window is already encoded in the equilibrium retarded self-energy $\Sigma_\mathbf{k}^R$ [Fig.~\ref{fig2}(a)] \cite{Mahan2000}.
The scattering rate $-\mathrm{Im}\,\Sigma^R_\mathbf{k}(\omega)$ is strongly suppressed for $\omega \in \mathcal{W}$, increasingly so at lower temperatures, because electron-like quasiparticles in this range can emit phonons only into states below the Fermi level, which are occupied, whereas hole-like quasiparticles can absorb phonons only by moving to states above the Fermi level, which are empty.
The corresponding quasiparticles are therefore long-lived, and relaxation inside $\mathcal{W}$ is strongly inhibited for lack of channels to dissipate the excess energy in quanta of $\Omega_\mathrm{r}$.
Note that while for weak excitations $\Sigma_\mathbf{k}^R$ accurately describes the return to equilibrium \cite{Sentef2013}, more generally, the nonthermal distributions matter as well, and in nonequilibrium the bottleneck is best viewed as a phase-space restriction on the energy 
emitted by
the electrons rather than as a single-particle lifetime effect \cite{Kemper2018}.

A phenomenon that cannot be explained only through suppressed single-particle scattering is evident in Video~\ref{vid1}, 
where one can discern a reduced phonon-energy window $\mathcal{W}_{1/2}=[-\Omega_\mathrm{r}/2,\Omega_\mathrm{r}/2]$. 
Its appearance is natural once the nonthermal fermion population is taken into account.
As the excited electrons relax toward the Fermi level, they start to fill the energy window $[0,\Omega_\mathrm{r}]$ from above (see Video~\ref{vid1}(b)). Due to the particle-hole symmetry of model \eqref{eq:holstein}, an analogous dynamics on the hole side leads to a hole accumulation in the window $[-\Omega_\mathrm{r},0]$ from below. The presence of excess electrons and holes in these windows activates relaxation processes within $\mathcal{W}$ which were previously blocked: electrons from the highly populated high-energy interval $[\Omega_\mathrm{r}/2,\Omega_\mathrm{r}]$ annihilate (under phonon emission) the holes in the more sparsely populated low-energy interval $[-\Omega_\mathrm{r}/2,0]$ and vice versa. The result of this annihilation process is an underpopulation of electrons and holes in $\mathcal{W}_{1/2}$, relative to the thermalized state (blue shaded region in Video~\ref{vid1}(a,b)). Because of the effectively cold distribution in $\mathcal{W}_{1/2}$, the remaining electrons and holes in $\mathcal{W}\setminus\mathcal{W}_{1/2}$ find no partners to recombine with, which results in the long-lived excess populations indicated by the red color in Video~\ref{vid1}(a,b).

In addition to the above, at early times ($t\lesssim 25$), we also resolve transient higher-order windows $\mathcal{W}_n=[-n\Omega_\mathrm{r},n\Omega_\mathrm{r}]$ with integer $n>1$, 
inside which the electron/hole relaxation is delayed for a short time, due to the inability of relaxation through phonon emission (occupied/unoccupied final states).  
We further recognize the oscillations of fermion-density deviation visible in Video~\ref{vid1}(b) and (d) as the behavior analyzed in Ref.~\cite{Kemper2013}, where oscillation frequencies set by $\epsilon_\mathbf{k}$, $\Omega_\mathbf{q}$, and the band edges were proposed as a means to experimentally map out the dispersion of unoccupied states.

Let us now discuss the \emph{nonequilibrium} changes to the retarded self-energy $\Sigma^R_\mathbf{k}(\omega,t)$ in the case of optical phonons. 
In Figs.~\ref{fig2}(b)-(d), we show the time-dependent deviation from the hot thermalized state, $-\mathrm{Im}\,\Delta\Sigma^R_\mathbf{k}(\omega,t)\equiv-[\mathrm{Im}\,\Sigma^R_\mathbf{k}(\omega,t)-\mathrm{Im}\,\Sigma^R_\mathbf{k}(\omega,t\to\infty)]$, with $-\mathrm{Im}\,\Sigma^R_\mathbf{k}(\omega,t)$ bearing the meaning of the scattering rate at $(\mathbf{k},\omega)$ and time $t$.
At early times, the self-energy exhibits a nontrivial $\mathbf{k}$ dependence.
These weak transient features slowly oscillate with frequency $\omega \sim 0.137$, leading to a periodic enhancement and weakening of the scattering particularly within the reduced phonon-energy window $\mathcal{W}_{1/2}$.
These features are strongly damped and fade at later times, which leaves an almost $\mathbf{k}$-independent self-energy, as in equilibrium [Fig.~\ref{fig2}(a)], with only faint features near $\Gamma$ and $M$ from phonon renormalization.
This near $\mathbf{k}$ independence of the long-time electron self-energy justifies, at this coupling and beyond transient effects, the local DMFT-like approximation used for it in previous works on the Holstein model \cite{Sentef2013,Werner2013,Kemper2014,Murakami2015,Rameau2016,Abdurazakov2018,Kemper2018}.
Note, however, that such an approximation cannot properly capture phonon renormalization, because the phonon self-energy is strongly momentum dependent even for weak $g$ (see the next section).
At larger $g$, where the electron renormalization gets stronger, the DMFT approximation would become unreliable for the electronic self-energy as well.

Apart from the transient effects, we notice a long-lived change in the self-energy near $\Omega_\mathrm{r}$ [Fig.~\ref{fig2}(d)].
Scattering is enhanced for $\omega>\Omega_\mathrm{r}$ and suppressed for $\Omega_\mathrm{r}/2<\omega<\Omega_\mathrm{r}$ (while inside $\mathcal{W}_{1/2}$ it is already essentially as weak as in equilibrium).
This agrees with the picture inferred from the electron population dynamics.
The reduced scattering within $\mathcal{W}\setminus\mathcal{W}_{1/2}$ reflects the effectively cold distribution within $\mathcal{W}_{1/2}$, which implies that the electrons and holes find no partners to recombine under phonon emission. At the same time, the deficit of electrons inside $[0,\Omega_\mathrm{r}/2]$, relative to the thermalized state, opens additional relaxation channels just above $\Omega_\mathrm{r}$ and enhances the scattering compared to equilibrium. The same is true for holes below $-\Omega_\mathrm{r}$, which can relax thanks to the deficit of holes in $[-\Omega_\mathrm{r}/2,0]$.
This nonequilibrium change of the scattering rate differs from that reported in Ref.~\cite{Kemper2014}. Since we treat renormalized phonons, the system thermalizes to a state hotter than the initial one instead of relaxing back to the initial state, and we measure the scattering change relative to the hot final state.

\begin{figure}[tb]
	\centering
	\includegraphics[width=\columnwidth]{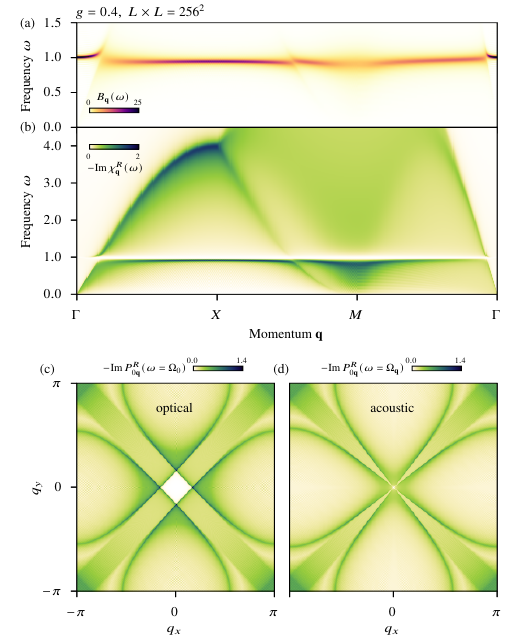}
	\caption{The correspondence between the renormalized phonon and charge spectra.
	(a) repeats the optical-phonon spectral function $B_\mathbf{q}(\omega)$ from Fig.~\ref{fig1}(a) and (b) shows the corresponding retarded charge susceptibility $-\mathrm{Im}\,\chi^R_\mathbf{q}(\omega)$.
	The contour-ordered susceptibility is defined as $\chi_\mathbf{q} = P_\mathbf{q}/(1 - P_\mathbf{q}\, V_{0\mathbf{q}})$, with $V_{0\mathbf{q}} = g_\mathbf{q}^2 D_{0\mathbf{q}}$ the bare phonon-mediated interaction.
	(c) shows the imaginary part of the bare particle-hole polarization $-\mathrm{Im}\,P^R_{0\mathbf{q}}(\omega)$ over the Brillouin zone at the optical phonon frequency $\Omega_0$, and (d) at the acoustic phonon frequency $\Omega_\mathbf{q}$.
	}
	\label{fig3}
\end{figure}

\subsubsection{Phonon relaxation (optical phonons)}\label{optical_ph_relaxation}

To set the stage for the discussion of the relaxation of the phonon densities, we note that the momentum-dependent self-energy $\Pi_\mathbf{q}$ renders the renormalized spectrum of an optical phonon dispersive, with a minimum at $\mathbf{q}=(\pi,\pi)$ and a maximum near $\mathbf{q}=(0,0)$ [Fig.~\ref{fig3}(a), which replots Fig.~\ref{fig1}(a)].
The renormalization is strongest at the nesting vector $(\pi,\pi)$, where it manifests as a decoherence of the phonon that turns into a softening at larger e-ph coupling, reflecting the charge-density-wave (CDW) instability of the system.
This behavior originates in the linear e-ph coupling, which ties the phonon propagator $D_\mathbf{q}$ to the charge susceptibility $\chi_\mathbf{q}$ through the relation $D_\mathbf{q}=D_{0\mathbf{q}}+g_\mathbf{q}^2\,D_{0\mathbf{q}}\,\chi_\mathbf{q}\,D_{0\mathbf{q}}$ \cite{Mahan2000,Weber2015}; see also Eqs.~\eqref{eq:pi}, \eqref{eq:P} and \eqref{eq:dysonD}.
The same features can thus appear in the $D_\mathbf{q}$ and $\chi_\mathbf{q}$ spectra, as they share the same poles and the same continuum, although the weights (controlled by the matrix elements of the displacement and density operators, respectively) are different.
The high momentum and frequency resolution of our analytic-continuation-free spectra makes this correspondence directly visible.
Comparing the phonon spectral function in Fig.~\ref{fig3}(a) with the retarded charge response $-\mathrm{Im}\,\chi^R_\mathbf{q}(\omega)$ in Fig.~\ref{fig3}(b) reveals shared structures.
Away from $(\pi,\pi)$, the phonon dispersion crosses the particle-hole continuum, and charge and phonon excitations hybridize.
This leads to anticrossings at the continuum edges, where the charge spectral weight $-\mathrm{Im}\,\chi^R_\mathbf{q}(\omega)$ accumulates.
This hybridization is long established, from the early theoretical analyses~\cite{Engelsberg1963,Engelsberg1964} to the phonon spectra of one-dimensional Holstein models~\cite{Sykora2006,Assaad2008,Hohenadler2011,Weber2015,Walther2026}, where the weight of the particle-hole continuum and the associated anticrossings have been resolved by a variety of methods.
Corresponding features are also present in two-dimensional equilibrium results~\cite{Li2015,Nowadnick2015,Dee2019,Nosarzewski2021}, but the lower resolution of those spectra hampers a direct comparison between the phonon and charge response.

We now turn to the relaxation of the phonon density $n^\mathrm{ph}_\mathbf{q}(t)$, whose momentum-resolved dynamics has not been analyzed within a full two-time NEGF treatment.
Panels (e) and (f) of Video~\ref{vid1} show the deviation $n^{\mathrm{ph}}_\mathbf{q}(t)-n^{\mathrm{ph}}_\mathbf{q}(\infty)$ for the optical-phonon model.
As discussed above, the cold initial state contains essentially no thermally excited phonons, whereas the final thermalized state is hotter.
The phonon density thus starts with a deficit throughout the BZ.
The phonon modes are subsequently populated across the entire BZ by the broadband excitation that the quench generates, as in the case of the electrons.
At early times, the deviation $n^{\mathrm{ph}}_\mathbf{q}(t)-n^{\mathrm{ph}}_\mathbf{q}(\infty)$ oscillates in sign, most clearly seen when playing Video~\ref{vid1} \cite{Vid1Supp}, and these oscillations propagate as ``waves'' across the Brillouin zone (similar waves are also visible in the early-time deviations of the electronic density).
By $t\simeq 10$, this reshuffling of energy between the electron and phonon subsystems results in an excess phonon population, relative to the thermalized final state, at most momenta.
The subsequent thermalization therefore requires a reabsorption of excess phonons, and this process remains slow up to the longest simulated times.

One furthermore recognizes sharp structures in the momentum dependence of the phonon density, around which the early-time oscillations take place.
These structures are particularly visible for $t\gtrsim 20$, most prominently as a diamond shape centered at $\mathbf{q}=(0,0)$.
They look nontrivial, but directly reflect the coupling between the phonon and charge spectra discussed above.
The momentum structure of the charge response is imprinted on the renormalized phonon dispersion and thereby on the phonon densities.
To see this, compare the snapshot in Video~\ref{vid1}(e) with Fig.~\ref{fig3}(c), which plots the imaginary part of the Lindhard charge response (i.e., the particle-hole polarization evaluated with the bare $G_{0\mathbf{k}}$) at the optical phonon frequency, $-\mathrm{Im}\,P^R_{0\mathbf{q}}(\omega=\Omega_0)$.
All the sharp features match, including the diamond centered at $(0,0)$, inside which the charge response vanishes as this region lies outside the particle-hole continuum.
Moreover, the phonon density relaxes faster at momenta where the charge response carries large weight (producing the bright features in Video~\ref{vid1}(e)), while inside the central diamond the deviations persist to the longest times.
In Video~\ref{vid1}(f), this diamond feature manifests itself as a highly populated band near $\mathbf{q}=(0,0)$.
The slow relaxation can be attributed to the weak self-energy effects, i.e., the weak renormalization of the phonon propagator at momenta where the charge response is small.
Long after the quench, these weakly renormalized modes are essentially described by the bare propagator $D_{0\mathbf{q}}$, meaning their occupation, once set, does not evolve further.
It is set during the early post-quench transient, when the evolving charge response couples the modes to electronic excitations, depositing an excess population.
As the transient settles and this coupling subsides, the modes freeze.
Hence, the excess phonons cannot be quickly reabsorbed in these momentum regions, which slows down the thermalization.
Around $\mathbf{q}=(\pi,\pi)$, by contrast, we observe a long-lived deficit of phonons, likely due to the phonon softening and decoherence at this wave vector [Fig.~\ref{fig3}(a)].
The system therefore not only exhibits a bottleneck in the electron relaxation, but also a separate one in the phonon thermalization.
While the former exists also for unrenormalized phonons, the latter is rooted in the momentum-dependent coupling to the particle-hole continuum and constitutes a distinct effect.

\subsection{Results for the model with acoustic phonons}

\subsubsection{Fermion relaxation (acoustic phonons)}\label{acoustic_fermion_relaxation}
We now turn to the fermion relaxation in the model with acoustic phonons [Video~\ref{vid1}(c) and~(d)].
Even though acoustic phonon modes can have arbitrarily small energies $\Omega_\mathbf{q}\to 0$, which in principle should allow electrons at any energy to relax and thus eliminate the relaxation bottleneck, this is not what we observe.
Instead, a clear bottleneck persists, with a nonthermal fermion population surviving in a momentum-energy window enclosing the Fermi surface.
The momentum dependence of $n^{\mathrm{e,h}}_\mathbf{k}(t)-n^{\mathrm{e,h}}_\mathbf{k}(\infty)$ is, however, richer than in the optical phonon case.
In particular, two characteristic energy scales define the window boundaries: $\Omega_0$ and $\Omega_0/\sqrt{2}$ (we find a slightly better fit to the windows using the bare, rather than the renormalized, frequency).
Moreover, an asymmetry emerges between the regions of the BZ that host an excess or a deficit of fermions relative to the thermalized state.

Similarly as for the optical phonon case, the properties of the phonon-energy window in the acoustic model can be understood by analyzing the equilibrium retarded self-energy $\Sigma^R_\mathbf{k}$.
Its imaginary part is shown in Fig.~\ref{fig2}(e).
As expected, for strongly dispersive phonons, the self-energy exhibits a clear momentum dependence; treating such a system thus clearly lies beyond the reach of DMFT approximations.
Nonetheless, there is a similarity between the acoustic [Fig.~\ref{fig2}(e)] and optical [Fig.~\ref{fig2}(a)] cases.
In the acoustic case, there likewise exists an energy range where the scattering rate $-\mathrm{Im}\,\Sigma^R_\mathbf{k}(\omega)$ is strongly suppressed, i.e., a phonon-energy window.
Two important differences stand out, however.
First, the sharp edge that defines the window is momentum dependent.
Second, there is weak spectral weight inside the window, centered around the Fermi momenta [which along the plotted high-symmetry path $\Gamma$-$X$-$M$-$\Gamma$ correspond to the points $X=(\pi,0)$ and midway between $M$ and $\Gamma$, i.e., $(\frac{\pi}{2},\frac{\pi}{2})$].
At $(\frac{\pi}{2},\frac{\pi}{2})$ this weight is very faint but nonzero, whereas near $X$ it is more pronounced.

To understand these features, one must recall that in the acoustic case scattering is possible only if both energy and momentum conservation are simultaneously satisfied.
This becomes explicit in the expression for the imaginary part of the equilibrium one-shot retarded self-energy~\cite{Mahan2000}
\begin{equation}
	\begin{aligned}
		-\mathrm{Im}\Sigma^R_\mathbf{k}(\omega)
		&=\\ \frac{1}{N_\mathbf{k}}\sum_\mathbf{q} g_\mathbf{q}^2
		\Bigl\{
			&\bigl[b_\beta(\Omega_\mathbf{q})+1-f_\beta(\epsilon_{\mathbf{k}+\mathbf{q}})\bigr]
			\delta\bigl(\omega-\epsilon_{\mathbf{k}+\mathbf{q}}-\Omega_{-\mathbf{q}}\bigr)
		\\
		\quad\phantom{\times\Bigl\{}
			&+\bigl[b_\beta(\Omega_\mathbf{q})+f_\beta(\epsilon_{\mathbf{k}+\mathbf{q}})\bigr]
			\delta\bigl(\omega-\epsilon_{\mathbf{k}+\mathbf{q}}+\Omega_\mathbf{q}\bigr)
		\Bigr\}.
	\end{aligned}\label{eq:imSigma}
\end{equation}
This expression explains why the in-window weight appears near the Fermi momenta. Only there does phase space exist for scattering with small momentum transfer $\mathbf{q}$, i.e., for processes in which empty and occupied states are separated by a small energy $\omega\to 0$ and a small momentum $\mathbf{q}\to 0$, allowing scattering via long-wavelength phonons.
The weight of this contribution remains small, largely because the e-ph coupling vanishes as $\Omega_\mathbf{q}\to 0$, which underlies the emergence of the phonon-energy window.

To determine the momentum dependence of the spectral features, consider a phonon emission process that scatters an electron from an arbitrary momentum $\mathbf{k}$ onto a specific final state $\mathbf{k}+\mathbf{q}=\mathbf{K}$.
From Eq.~\eqref{eq:imSigma}, the contribution to $-\mathrm{Im}\Sigma^R_\mathbf{k}$ appears at $\omega_{\mathbf{k}\to\mathbf{K}} = \epsilon_{\mathbf{K}} + \Omega_{\mathbf{K}-\mathbf{k}}$ (using $\Omega_{-\mathbf{q}}=\Omega_{\mathbf{q}}$).
In Fig.~\ref{fig2}(e), we show the quantity $\omega_{\mathbf{k}\to X,(\frac{\pi}{2},\frac{\pi}{2})}$ for scattering onto the Fermi surface points along the high-symmetry path $\Gamma$-$X$-$M$-$\Gamma$. The black solid line marks scattering onto $X$, and the black dashed line marks scattering onto $(\frac{\pi}{2},\frac{\pi}{2})$.
These curves clearly delimit the in-window contributions.
Note that the spectral weight near $X$ is larger than near $(\frac{\pi}{2},\frac{\pi}{2})$ because $X$ is a van Hove point of the two-dimensional square lattice, where the density of states is enhanced.
The momentum-dependent edge of the acoustic phonon-energy window arises from the strongest contribution to $-\mathrm{Im}\Sigma^R_\mathbf{k}$, which comes from scattering to a van Hove point.
To illustrate this, consider the final state $\mathbf{K}=X'=(0,\pi)$ (chosen so that we also see scattering between the van Hove points, i.e., $X\to X'$).
The corresponding $\omega_{\mathbf{k}\to X'}$ is shown as a dashed white curve in Fig.~\ref{fig2}(e) and clearly marks the acoustic phonon-energy window.

Equipped with this understanding of the equilibrium self-energy, we can discuss the relaxation of the fermion densities for acoustic phonon modes.
In Video~\ref{vid1}(c) and~(d), we mark contours at $\pm\omega_{\Gamma\to X'}=\pm\omega_{M\to X'}=\pm\Omega_0/\sqrt{2}$ and $\pm\omega_{X\to X'}=\pm\Omega_0$, which are the extremal values of the boundary $\omega_{\mathbf{k}\to X'}$ of the acoustic phonon-energy window.
As expected, around the van Hove points $X$ and $X'$ the nonthermal fermion population persists within $[-\Omega_0,\Omega_0]$, while around $(\frac{\pi}{2},\frac{\pi}{2})$ it persists within $[-\Omega_0/\sqrt{2},\Omega_0/\sqrt{2}]$ (see the main panel in Video~\ref{vid1}(c)).
Furthermore, the reduced phonon-energy window $\mathcal{W}_{1/2}$ identified in the optical case is present here as well, and its boundaries mark the separation between the red regions with an excess of fermions and the blue regions with a fermion deficit.

The spatial pattern of excess and deficit reveals additional structure compared to the optical phonon case.
The excess fermion weight (red) around $(\frac{\pi}{2},\frac{\pi}{2})$ is larger than around $X=(\pi,0)$.
Furthermore, at longer times [$t\approx 175$; see the inset in Video~\ref{vid1}(d) and~\cite{Vid1Supp} to play the video], there is no excess of fermions around $X$ anymore, only around $(\frac{\pi}{2},\frac{\pi}{2})$, and no deficit of fermions around $(\frac{\pi}{2},\frac{\pi}{2})$, only around $X$.
This asymmetry can be understood as follows.
The excess fermions with energies within $[\Omega_0/\sqrt{2},\Omega_0]$ around $X$ can scatter rather efficiently to fill the deficit around $(\frac{\pi}{2},\frac{\pi}{2})$, because they have the necessary energy of at least $\omega_{X\to(\frac{\pi}{2},\frac{\pi}{2})}=\Omega_0/\sqrt{2}$ [see Fig.~\ref{fig2}(e) for the $\omega_{\mathbf{k}\to(\frac{\pi}{2},\frac{\pi}{2})}$ shown as dashed black curve, and see Fig.~\ref{fig1}(c) for $\Omega_0/\sqrt{2}$ marked on the Brillouin zone].
The reverse process is suppressed, because the excess fermions around $(\frac{\pi}{2},\frac{\pi}{2})$ are trapped in the window with energy $\omega<\Omega_0/\sqrt{2}$, 
insufficient to scatter into the deficit region around $X$.
Thus, excess fermions from the vicinity of $X$ fill the deficit near $(\frac{\pi}{2},\frac{\pi}{2})$.
As a result, at long times a deficit persists around $X$, which cannot be filled by $\mathbf{q}\to 0$ processes because there is no sufficient excess fermion population nearby.
Conversely, around $(\frac{\pi}{2},\frac{\pi}{2})$ an excess survives, which cannot relax because the nearby deficit has already been filled by scattering from around $X$ (together with weak $\mathbf{q}\to 0$ processes).
An asymmetry thus emerges in the Brillouin zone between the locations of excess and deficit fermion populations.
We conclude that the acoustic bottleneck in the fermion relaxation persists both because low-$\mathbf{q}$ scattering is weak (owing to the vanishing e-ph coupling at small $\mathbf{q}$) and because of the asymmetry that develops during the dynamics.
Let us also note that, in contrast to the optical case, there is no clear evidence for an early-time relaxation through a cascade of higher-order windows $\mathcal{W}_n$ with $n>1$.

We finally comment on the nonequilibrium changes to the retarded self-energy $\Sigma_\mathbf{k}^R(\omega,t)$ for acoustic phonons.
The main features are analogous to those in the optical case.
Scattering is enhanced just above the phonon-energy window edge ($\omega > \omega_{\mathbf{k}\to X'}$) and suppressed just below it.
Inside the window, the scattering is decreased relative to the thermalized state near the Fermi surface (i.e., for the in-window contribution with low momentum $\mathbf{q}$ transfer), 
and increased elsewhere in the Brillouin zone [Figs.~\ref{fig2}(g) and~(h)].
This is consistent with the asymmetric distribution of fermion excess and deficit, which reduces the phase space for scattering with low-momentum phonons.
We also see a faint blue shadow of decreased scattering that traces the shape of the electronic dispersion.
Unlike the optical case, we find no slow oscillations in the self-energy deviation.
The oscillations of the fermion-density deviation of the type discussed in connection with Ref.~\cite{Kemper2013} in Sec.~\ref{optical_fermion_relaxation} are nevertheless present in the acoustic phonon data of Video~\ref{vid1}(d).

\subsubsection{Phonon relaxation (acoustic phonons)}\label{acoustic_ph_relaxation}

We now turn to the phonon relaxation in the acoustic model.
Here, already the initial response to the quench differs from the optical case.
Whereas for optical phonons the quench triggers an overpopulation of phonons across the entire BZ, in the acoustic case the initial overpopulation is confined to momenta near $(\pi,\pi)$, where the phonon energy $\Omega_\mathbf{q}$ is largest [compare Video~\ref{vid1}(f) and~(h) at $t=2$, or play the video here~\cite{Vid1Supp}].
This occurs because, in the hot thermalized state, the low-$\mathbf{q}$ modes have a much higher occupation than in the cold initial state, and the brief post-quench dynamics is insufficient to populate them.
Instead, these low-$\mathbf{q}$ phonons are emitted slowly over the course of the subsequent evolution, gradually reducing the size of the deficit (blue) region; see Video~\ref{vid1}(h).
The shape of this deficit region follows the contours of the acoustic phonon dispersion $\Omega_\mathbf{q}$ at successively lower energies $\omega$; at long times [$t\gtrsim 68$, Video~\ref{vid1}(g)] it takes a roughly circular form, reflecting the conical dispersion at $\mathbf{q}\to 0$.
The emission of low-$\mathbf{q}$ phonons is slow both because the e-ph coupling vanishes as $\mathbf{q}\to 0$ and because of the fermion asymmetry discussed in the previous subsection, which restricts the available phase space for scattering.
The acoustic model thus displays a distinct kind of phonon-relaxation bottleneck: a persistent difficulty in emitting low-$\mathbf{q}$ phonons and reaching their large thermal occupation in the hot final state.

Apart from this low-$\mathbf{q}$ bottleneck, the sharp momentum-dependent structures in the phonon density deviation that we identified for the optical-phonon model are also visible in the acoustic case [compare Video~\ref{vid1}(e) and~(g)].
The explanation is analogous.
The momentum structure of the charge response is imprinted on the renormalized phonon dispersion and thereby on the phonon densities.
To confirm this, we plot in Fig.~\ref{fig3}(d) the Lindhard charge response at the dispersive acoustic phonon frequency, $-\mathrm{Im}\,P^R_{0\mathbf{q}}(\omega=\Omega_\mathbf{q})$.
The agreement with the bright features in Video~\ref{vid1}(g) is clear.
These bright spots again mark the regions that thermalize fastest, since the stronger charge response leads to a larger phononic self-energy and hence more efficient (re)absorption processes for the overemitted phonons.
However, the reabsorption bottleneck that we noted for optical phonons is not prominent here.
The reason is that, unlike in the optical case, all acoustic phonon energies $\Omega_\mathbf{q}$ lie within the particle-hole continuum, so every phonon mode has a nonnegligible self-energy and is renormalized to some degree.
As a consequence there exist no phonon modes for which absorption processes are almost completely blocked, which was the situation that caused the persistent red diamond around $\Gamma$ in the optical-model data [Video~\ref{vid1}(e)].

\section{Discussion and conclusions}\label{conclusions}

In this work, we used the recently developed QTT-NEGF framework to investigate the momentum-resolved nonequilibrium dynamics of electron-phonon systems on lattices up to $256\times 256$ sites, comparing optical and acoustic phonon dispersions.
This unprecedented computational capability, enabled by the huge memory compression of the quantics-tensor-train representation of correlation functions, allowed us to study the relaxation dynamics over long propagation times with full momentum resolution and phonon dispersion, a task previously inaccessible to the diagrammatic NEGF approach employed here, which retains the full two-time dependence of the functions.
By analyzing the evolution of the electron, hole, and phonon densities, as well as the self-energies, we revealed a rich hierarchy of relaxation bottlenecks that substantially extends and refines the understanding of the phonon-window effect discussed in earlier works~\cite{Sentef2013,Kemper2014,Murakami2015,Rameau2016,Abdurazakov2018,Kemper2018}.

For optical phonons, we confirmed the existence of the main phonon-energy window $\mathcal{W}=[-\Omega_\mathrm{r},\Omega_\mathrm{r}]$, within which single-particle scattering is strongly suppressed and the energy transfer from the electron to the phonon subsystem is impeded. This mechanism leads to long-lived nonthermal electron populations within the window.
In particle-hole-symmetric photo-doped systems, our momentum-resolved analysis uncovered an additional reduced phonon-energy window $\mathcal{W}_{1/2}=[-\Omega_\mathrm{r}/2,\Omega_\mathrm{r}/2]$, inside which the fermion population remains effectively cold compared to the thermalized distribution up to the longest simulation times.
This feature emerges from the mutual annihilation of electrons and holes across the two halves of $\mathcal{W}_{1/2}$ under phonon emission, which results in a long-lived deficit within the reduced window and the accumulation of an excess population in $\mathcal{W}\setminus\mathcal{W}_{1/2}$.
At early times ($t\lesssim 25$), we also resolved transient higher-order windows $\mathcal{W}_n=[-n\Omega_\mathrm{r},n\Omega_\mathrm{r}]$ with $n>1$, where the fermion relaxation is briefly delayed by short-lived phase-space restrictions.
This is reminiscent of the cascade optical phonon emission studied in graphene \cite{Ryzhii2007,Vasko2010,Kim2011}.
We further resolved a separate bottleneck in the phonon thermalization for optical modes, rooted in the momentum-dependent coupling of the phonon propagator to the particle-hole continuum.
Phonon modes at momenta where the charge response is strongly suppressed remain essentially unrenormalized and cannot efficiently reabsorb the excess phonons emitted in the early-stage dynamics by the relaxing electrons, leaving a persistent phonon overpopulation in those regions of the Brillouin zone.

Previous studies of electron-phonon bottlenecks~\cite{Sentef2013,Kemper2014,Rameau2016,Kemper2018,Abdurazakov2018,Murakami2015} did not explicitly discuss the reduced window, although the underlying reasons differ.
In Refs.~\cite{Sentef2013,Kemper2014,Rameau2016,Kemper2018}, the phonons were treated as unrenormalized, so that the system relaxed back to the cold initial state after dissipating the excitation energy.
When the electron-density deviation is measured relative to this cold reference, one finds a fermion excess throughout the Brillouin zone, since the transient hot distribution is broader than the cold state into which the fermions relax. This masks the signed pattern of excess and deficit that reveals $\mathcal{W}_{1/2}$.
By measuring deviations from the hot thermalized final state, our analysis instead exposes both regions directly.
Reference~\cite{Abdurazakov2018} treated renormalized phonons, yet the reduced window was not pointed out there, possibly because of the absence of particle-hole symmetry in the model (as also in the experimental study \cite{Rameau2016}), and a smaller phonon frequency that narrows $\mathcal{W}_{1/2}$.
In contrast, Ref.~\cite{Murakami2015} employed a coupling quench similar to ours, a particle-hole-symmetric model, relatively large phonon frequencies, and renormalized phonons within DMFT, and a signature of $\mathcal{W}_{1/2}$ is indeed clearly visible in their data; it was simply not explicitly discussed.
This is presumably because the feature is subtle when viewed through the electron distribution alone, whereas plotting the sign of the momentum-resolved deviation from the thermalized state, as we do in Video~\ref{vid1}, reveals it unambiguously.

For acoustic phonons, the situation is considerably richer.
We emphasize that acoustic phonons have not previously been studied with a reliable diagrammatic NEGF approach, because the memory demand is prohibitive without the QTT compression.
Here, the phonon-energy window $\mathcal{W}$ persists, even though acoustic phonons extend to arbitrarily small energies and could in principle relax carriers in small steps. It acquires a pronounced momentum dependence, with its width varying across the Brillouin zone as dictated by simultaneous energy and momentum conservation.
The reduced window $\mathcal{W}_{1/2}$ is present as well, but it becomes asymmetric: the locations of excess and deficit fermion populations are separated in momentum space because scattering between different regions is directional.
Fermions with sufficient energy can scatter to fill a deficit in another region, while those with lower energy remain trapped.
This asymmetry, combined with the vanishing electron-phonon coupling at $\mathbf{q}\rightarrow 0$, creates a particularly persistent bottleneck for low-momentum phonon modes.
Namely, the thermal occupation of acoustic phonons near $\mathbf{q}=0$ in the hot final state is large, yet these modes are emitted only very slowly, resulting in a long-lived phonon deficit at the zone center.
This inefficient low-$\mathbf{q}$ emission leads to the emergence of the windows $\mathcal{W}$ and $\mathcal{W}_{1/2}$ with nonthermal fermion populations in the acoustic-phonon case.
We also showed that the low-$\mathbf{q}$ bottleneck coexists with the charge-response-induced structures identified in the optical case, although the reabsorption bottleneck is less pronounced for acoustic phonons because every mode lies within the particle-hole continuum and experiences some degree of renormalization.
In both the optical and acoustic cases, the high resolution of our spectra makes the correspondence between the phonon and charge response directly visible.
While this correspondence was established in equilibrium~\cite{Engelsberg1963,Engelsberg1964,Sykora2006,Assaad2008,Hohenadler2011,Weber2015,Walther2026}, we showed it also governs the nonequilibrium relaxation, with the momentum structure of the particle-hole continuum deciding which phonon modes thermalize quickly and which remain frozen.

We note that acoustic-phonon features similar to those described above appear in a few earlier studies.
Overheating of high-energy acoustic phonons and slow relaxation of low-energy modes was previously reported in a Boltzmann-equation study~\cite{Held2025}, which arrived at this result within a simpler treatment that assumed thermal electrons at all times, thus missing the asymmetry in fermion populations.
Furthermore, in a development separate from the phonon-window studies of Refs.~\cite{Sentef2013,Kemper2014,Rameau2016,Kemper2018,Abdurazakov2018,Murakami2015}, the existence of the main optical-phonon-energy window and the impeded fermion relaxation within it due to slow emission of low-$\mathbf{q}$ acoustic phonons was recognized in graphene as a route to population inversion and THz lasing: Cascade optical-phonon emission first accumulates carriers below the optical frequency, where only the inefficient acoustic-phonon scattering can relax them, leading to long-lived population inversions under steady-state driving~\cite{Ryzhii2007,Vasko2008,Vasko2010,Kim2011}.
Our work suggests that comparable long-lived population inversions can arise from relaxation with purely acoustic phonons.

The analysis of the retarded electron self-energy $\Sigma^R_\mathbf{k}(\omega,t)$ corroborates our conclusions and provides additional insights.
For optical phonons, the deviations of $\Sigma^R_\mathbf{k}$ from equilibrium are only weakly $\mathbf{k}$-dependent at long times, which retrospectively justifies the local electron self-energy approximations employed in earlier DMFT-based studies~\cite{Sentef2013,Werner2013,Kemper2014,Murakami2015,Rameau2016,Abdurazakov2018,Kemper2018}.
However, even in the optical case and for weak $g$, the phonon self-energy remains highly momentum dependent as it is tied to the charge susceptibility. 
A local approximation is thus insufficient to properly describe the phonon renormalization and, consequently, the bottleneck in the phonon-density relaxation.
For acoustic phonons, $\Sigma^R_\mathbf{k}$ remains strongly $\mathbf{k}$-dependent at all times, and, as expected, a fully momentum-resolved treatment is essential for dispersive phonons.
The nonequilibrium changes to the scattering rate near the window edges are consistent with the asymmetric fermion distributions and the phase-space restrictions they impose.

The methodology demonstrated here opens several avenues for future investigations.
The QTT-NEGF framework can be extended to treat realistic phonon dispersions obtained from first-principles calculations, enabling material-specific simulations of nonequilibrium dynamics at system sizes which are large enough to eliminate artificial recurrences.
The approach is also naturally suited to study the interplay of electron-phonon coupling with other interactions, such as the Hubbard $U$, or to investigate the transient enhancement of charge order susceptibilities that we observed under certain quenching protocols.
The efficiency of the method demonstrated here is promising: further algorithmic improvements should enable still larger system sizes and longer propagation times. Even without such advances, it is already possible to treat multiple renormalized phonon branches---for instance, one acoustic and one optical mode simultaneously---and to study their interplay.

Our results have direct implications for the interpretation of time-resolved spectroscopies.
The momentum-dependent electron densities that we computed correspond, through the lesser Green's function $\mathrm{Im}\, G^<_\mathbf{k}(\omega,t)$, to the spectral intensity measured in time-resolved angle-resolved photoemission spectroscopy (tr-ARPES).
The relaxation bottlenecks and the asymmetric fermion distributions that we identified should therefore be observable in tr-ARPES experiments. The most promising targets are materials with strong electron-phonon coupling and large phonon frequencies, such as molecular crystals \cite{Nomura2016,Buzzi2020}.
In particular, the reduced phonon-energy window $\mathcal{W}_{1/2}$ manifests as a region of the Brillouin zone where the spectral weight remains below its thermalized value long after the excitation, providing a clear experimental signature.
The long-lived nonequilibrium phonon populations that we predict could be detected through ultrafast electron diffuse scattering~\cite{deCotret2019,Otto2021,Seiler2021,Zacharias2021a,Zacharias2021b,Mo2024}.
More broadly, our work establishes the QTT-NEGF approach as a scalable and controlled simulation framework for the quantitative study of nonequilibrium electron-phonon dynamics, which overcomes the lattice-size and propagation-time limitations that have constrained diagrammatic treatments of this problem.

\begin{acknowledgments}
We thank M. Schüler for helpful discussions and acknowledge support from SNSF Grant No.\ 200021-196966.
The calculations were performed on the beo05 cluster at the University of Fribourg.
We used Cursor~3.8 with various large language model backends for scientific reasoning and interpretation, drafting and revising scientific claims, literature synthesis, proofreading and editing, and the generation of plotting scripts.
The artificial intelligence tools were directed through iterative prompting and all generated content was reviewed, edited, and verified by the authors.
\end{acknowledgments}

\bibliography{ref}

\end{document}